\documentclass[galaxies,article,submit,moreauthors,pdftex,10pt,a4paper,dvipsnames]{mdpi}

\firstpage{1}
\articlenumber{5}
\doinum{10.3390/galaxies8010005}
\pubvolume{8}
\pubyear{2020}
\copyrightyear{2020}
\history{Received: 12 November 2019; Accepted: 25 December 2019; Published: 30 December 2019}

\usepackage{physics}
\usepackage{tikz}
\usepackage{verbatim}
\usepackage{hyperref}
\usetikzlibrary{arrows,babel}
\usepackage[labelfont={bf},labelsep=endash]{caption}
\usetikzlibrary{positioning,arrows}
\usetikzlibrary{decorations.pathmorphing}
\usetikzlibrary{decorations.markings}
\tikzset{
momentum/.style={postaction={decorate},decoration={markings,mark=at position 1 with {\arrow{>}}}},
particle/.style={dashed
    },
photon/.style={decorate, 
    decoration={snake}},
    math/.style={draw, execute at begin node={$\displaystyle}, execute at end node={$}}
 }

\pdfoutput=1

\Title{Constraints to dark matter annihilation from high-latitude HAWC unidentified sources}

\Author{Javier Coronado-Bl\'azquez $^{1,2}$, Miguel A. S\'anchez-Conde $^{1,2}$}

\AuthorNames{Javier Coronado-Bl\'azquez, Miguel A- S\'anchez-Conde}

\address{%
$^{1}$ \quad Instituto de F\'{i}sica Te\'{o}rica UAM-CSIC,\\Universidad Aut\'{o}noma de Madrid, C/ Nicol\'{a}s Cabrera, 13-15, 28049 Madrid, Spain\\
$^{2}$ \quad Departamento de F\'{i}sica Te\'{o}rica, M-15,\\Universidad Aut\'{o}noma de Madrid, E-28049 Madrid, Spain}

\corres{Correspondence: javier.coronado@uam.es, miguel.sanchezconde@uam.es}

\abstract{The $\Lambda$CDM cosmological framework predicts the existence of thousands of subhalos in our own Galaxy not massive enough to retain baryons and become visible. Yet, some of them may outshine in gamma rays provided that the dark matter is made of weakly interacting massive particles (WIMPs), that would self-annihilate and would appear as unidentified gamma-ray sources (unIDs) in gamma-ray catalogs. Indeed, unIDs have proven to be competitive targets for dark matter searches with gamma rays. In this work, we focus on the three high-latitude ($\abs{b}\geq10Âº$) sources present in the 2HWC catalog of the High Altitude Water Cherenkov (HAWC) observatory with no clear associations at other wavelenghts. Indeed, only one of these sources, 2HWC J1040+308, is found to be above the HAWC detection threshold when considering 760 days of data, i.e., a factor 1.5 more exposure time than in the original 2HWC catalog. Other gamma-ray instruments such as Fermi-LAT or VERITAS at lower energies do not detect the source. Also, this unID is reported as spatially extended, making it even more interesting in a dark matter search context. While waiting for more data that may shed further light on the nature of this source, we set competitive upper limits on the annihilation cross section by comparing this HAWC unID to expectations based on state-of-the-art N-body cosmological simulations of the Galactic subhalo population. We find these constraints to be particularly competitive for heavy WIMPs, i.e., masses above $\sim 25$ (40) TeV in the case of the $b\bar{b}$ ($\tau^+\tau^-$) annihilation channel, reaching velocity-averaged cross section values of $2\cdot10^{-25}$ ($5\cdot10^{-25}$) $cm^3s^{-1}$. Although far from testing the thermal relic cross section value, the obtained limits are independent and nicely complementary to those from radically different DM analyses and targets, demonstrating once again the high potential of this DM search approach.}


\begin{document}

\section{Introduction}
As of today, we believe that about 85\% of all the matter in the Universe is of non-baryonic nature \cite{Planck15,GarrettDuda09,Freese09}. Despite its large abundance, the ultimate nature of this so-called dark matter (DM) remains unknown, being at present one of the most puzzling questions in modern physics.

N-body cosmological simulations reveal that DM structures form hierarchically in a bottom-up scenario: the DM particles first collapse into small gravitationally bound systems (known as halos), and then form more massive structures through a complex history of merging and accretion. As a result, DM halos contain a very large number of smaller {\it subhalos} \cite{Madau2008,Zavala2019}.

For DM candidates with weak-scale masses and interactions, as those preferred by supersymmetric particle physics theories \cite{Jungman1996}, subhalos with masses between approximately $10^{-11}-10^{-3}$ M\textsubscript{\(\odot\)} \cite{Bertschinger06, Profumo+06, Bringmann09} (depending on the specific DM particle model) up to roughly $10^{10}$ M\textsubscript{\(\odot\)} are expected to exist in a Milky Way-like galaxy. The vast majority of the Galactic DM subhalos are not expected to be massive enough to host baryons and therefore remain completely dark, with no visible counterparts \cite{Gao2004,Ocvirk2016,Sawala2015,Sawala2015a,Fitts2018}. Yet, many of these light subhalos or {\it dark satellites} will lie much closer to the Earth compared to the most massive ones, given both their much larger number density and higher survival probability at small Galactocentric radii against tidal disruption \cite{vlii_paper, Springel2008}.

If the DM is made of the so-called \textit{Weakly Interacting Massive Particles} (WIMPs) DM \cite{Roszkowski+17,Bertone10}, these small subhalos may outshine in gamma rays. WIMPs can achieve the correct relic DM abundance (the so-called "WIMP miracle"), through self-annihilation in the early Universe. This process gives rise to a Standard Model (SM) particle-antiparticle pair which, among other possible subsequent by-products, may yield gamma-ray photons \cite{Bertone+05, Bertone2009}. Small subhalos may be potentially interesting for this kind of {\it indirect} dark matter search since, as mentioned, many of them may be located close enough to Earth to yield high enough gamma-ray fluxes \cite{Anderson2010, Calore+17, Coronado_Blazquez2019, aguirre2019}. Interestingly, an important fraction of objects (typically between $30-40\%$, depending on the catalog) in the very high energy (VHE, $E>100$ GeV) sky are unidentified sources (unIDs), i.e., objects with no clear single association or counterpart. Some of these unIDs may actually be DM subhalos \cite{Bertoni+15, Bertoni+16, BuckleyHooper10, Calore+17, fermi_dm_satellites_paper, HooperWitte17, Egorov2018, Angelis2018, Zechlin+12, ZechlinHorns12, Coronado_Blazquez2019}. In our work, we will explore this possibility focusing on the unIDs recently detected in the VHE regime by the High Altitude Water Cherenkov Observatory (HAWC) \cite{Abeysekara2013,Abeysekara2019}. 

One of the greatest challenges when using dark satellites to search for DM is to come up with a reliable characterization of the low-mass subhalo population that would allow for robust predictions of their DM annihilation fluxes. Indeed, there is no N-body cosmological simulation at present able to resolve the entire Galactic subhalo population all the way down to the expected minimum subhalo mass. In this paper, we will adopt the results in \cite{aguirre2019}, where the authors devised an algorithm to repopulate the original Via Lactea II (VL-II) \cite{vlii_paper} N-body cosmological simulation with low-mass subhalos below its resolution limit, of about $\sim 10^5 \mathrm{M\textsubscript{\(\odot\)}}$. To do so, they first studied what found for the abundance and distribution of resolved subhalos in the simulation, and extrapolated the relevant quantities to smaller subhalo masses. In addition, they adopted state-of-the-art models to describe the subhalo structural properties \cite{Moline+17}. With the predictions of \cite{aguirre2019} at hand, and in the absence of an obvious DM subhalo candidate among the pool of HAWC unIDs, which is actually reduced to just one source in the end, 2HWC J1040+308, we will show in this work that it is possible to place competitive constraints on the WIMP DM parameter space following the procedure described in \cite{Calore+17,Coronado_Blazquez2019}.

The structure of this paper is as follows. In Section \ref{sec:hawc_bigsection} we briefly describe the observational status of the VHE sky, paying special attention to the HAWC observatory. We also address our specific target selection in this section. Section \ref{sec:constraints} is devoted to the computation of the DM constraints taking into account the different instrumental parameters and the theoretical predictions from N-body simulations. We conclude in Section \ref{sec:summary}, where we also make explicit the caveats in our analysis, compare our results to previous work and make qualitative assessments for future unID-based DM search strategies and experiments.

\section{HAWC unidentified sources}
\label{sec:hawc_bigsection}

In this section, we briefly review the HAWC observatory and discuss the target selection we made for this work.

\subsection{HAWC and the TeV sky}\label{sec:hawc}
The High Altitude Water Cherenkov Observatory (HAWC) is a VHE gamma-ray detector with a one-year sensitivity of $5-10\%$ of the flux of the Crab Nebula. Unlike imaging atmospheric Cherenkov telescopes (IACTs), such as H.E.S.S. \cite{Aharonian+2004}, MAGIC \cite{Aleksic+2016} or VERITAS \cite{Holder+2006}, which detect the Cherenkov light emitted by the extensive air showers produced by the gamma rays in the atmosphere, HAWC detects particles of these air showers that reach ground level with water-based Cherenkov detectors. Therefore, HAWC is able to operate continuously and to achieve a field of view (FoV) larger than 1.5 sr \cite{Smith2007}. HAWC is located on the flanks of the Sierra Negra volcano near Puebla, Mexico, and consists on a large pond of water tanks, each of 7.3m of diameter, located at 4100 m elevation. The pond contains 900 photomultiplier tubes (PMTs), while each tank contains 200,000 liters of purified water \cite{Smith2015}. Starting operations in 2012 with 30 tanks, they were progressively increased to 300 in 2015, detecting TeV gamma and cosmic rays with a high duty cycle (>95\%) \cite{Dingus2007,hawc_icrc2013}. 

\begin{figure}[H]
\centering
\includegraphics[height=6.3cm]{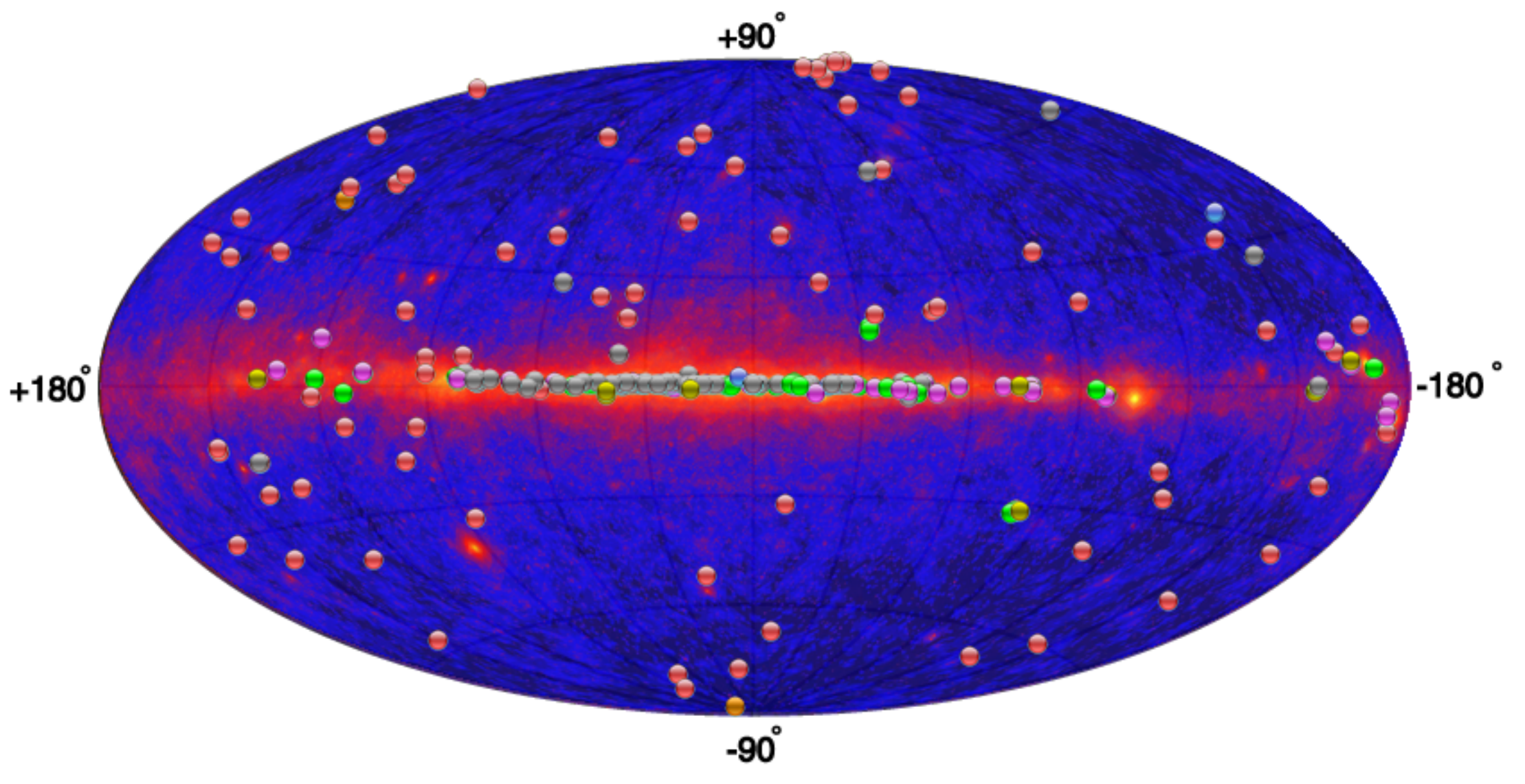}
\includegraphics[height=6.3cm]{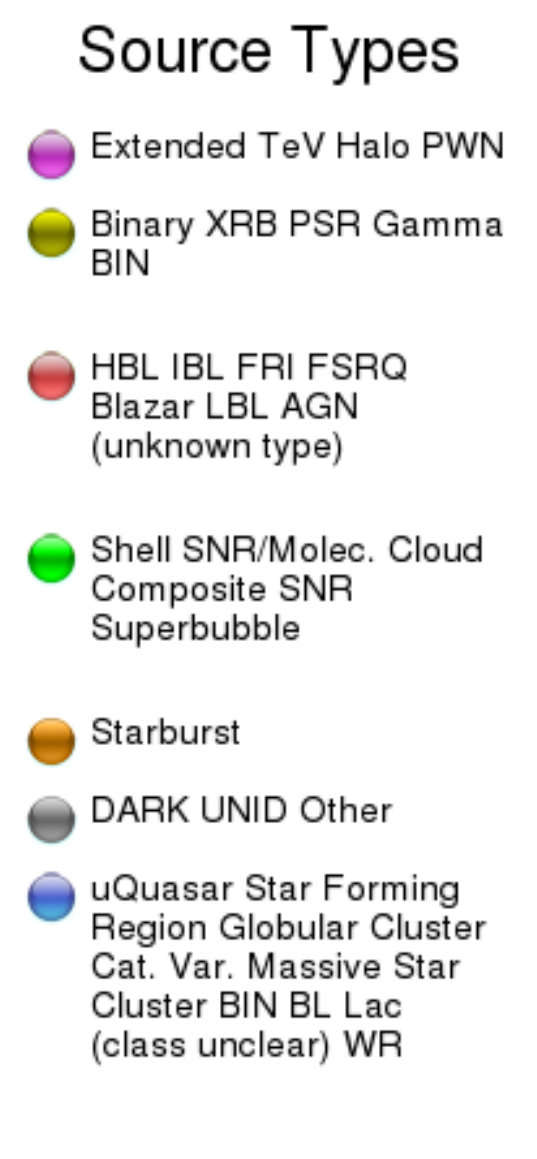}
\caption{VHE sources listed in the TeVCat online catalog as of November 2019, here shown in Hammer-Aitoff projection and Galactic coordinates. Background is the gamma-ray sky as seen by Fermi-LAT in 4 years of operation. Note the higher density of sources, in particular of unIDs, along the Galactic plane. The figure was generated with the TeVCat online tool \cite{WakelyHoran2008}.}
\label{fig:tevcat_map}
\end{figure}

Currently, there are about 200 known VHE gamma-ray sources detected by a handful of observatories, compiled within the TeVCat\footnote{\url{tevcat2.uchicago.edu}} catalog \cite{WakelyHoran2008}. A Galactic map in Hammer-Aitoff projection with all TeVCat sources and their class is plotted in figure \ref{fig:tevcat_map}. From these, roughly 60 sources remain with no clear association. Most of them are probably astrophysical sources of Galactic origin, since many were discovered after the Galactic plane survey by H.E.S.S \cite{hess_plane_survey}. Yet, there are several unIDs located at high Galactic latitudes, in particular three of them reported in the 2HWC catalog -- the second HAWC catalog \cite{2HWC_paper} comprising 507 days of instrument operation. In this work, we will focus on these sources.

\subsection{Target selection}\label{sec:target_selection}
We chose to work only with the high-latitude unIDs listed in the 2HWC catalog mainly for two reasons:

\begin{itemize}
    \item \textbf{Why HAWC?} - HAWC is currently the only VHE observatory able to survey a significant fraction of the whole sky. Indeed, the combination of its 1.5 sr FoV,  Earth rotation and location in Mexican soil, $19^\circ$ above the Equator, makes it possible to observe 2/3 of the sky every day.\footnote{In fact, HAWC applies a zenith cut of $45^\circ$, which for a daily exposure translates in 8.4sr, corresponding exactly to 66.85\% of the sky.} In contrast, IACTs are pointing telescopes, i.e., they need to know the location of the sources in advance and, thus, just point towards specific targets. For the purposes of this work, it becomes critical to have observations of a large fraction of the sky. This is so because we adopt the methodology introduced in \cite{Coronado_Blazquez2019} to set constraints on the DM annihilation parameter space using unIDs. The method is based on a comparison between the number of catalogued unID sources and subhalo predictions derived from N-body cosmological simulations. Large fractions of the sky, observed with a nearly uniform exposure, will allow for a more accurate comparison to N-body simulations, as we can derive a more robust statistical determination of the subhalo annihilation fluxes for significantly large sky areas. We remind that these fluxes are proportional to the so-called J-factor:
   
    \begin{equation}
    \label{eq:j_factor}
    J=\int_{\Delta\Omega}d\Omega\int_{l.o.s} \rho_{DM}^2\left[r\left(\lambda\right)\right]d\lambda,
    \end{equation}
    
    \noindent where the first integral is performed over the solid angle of observation ($\Delta\Omega$), the second one along the line of sight (l.o.s, $\lambda$), and $\rho_{DM}$ is the dark matter density profile of a given subhalo. 
    \\
    \item \textbf{Why high latitude?} $\Lambda$CDM predicts the existence of DM subhalos at all Galactic latitudes distributed nearly isotropically from the Sun's position in the Galaxy. On the other hand, the majority of Galactic VHE astrophysical sources, such as pulsars, binaries, supernova remnants and pulsar wind nebulae, are expected to cluster heavily along the Galactic plane, where most of the stars and gas reside. Since many of these objects are expected to also be hidden among the pool of unIDs awaiting for a proper classification, we expect the distribution of unIDs to peak around zero Galactic latitude as well (as already discussed in Figure \ref{fig:tevcat_map}). For our purposes, these low-latitude sources only add contamination to our sample of potential DM subhalo candidates. Therefore, we apply a cut in our analysis at Galactic latitudes $|b|\leq 10^\circ$. For consistency, this cut will also need to be done on our predicted subhalo distribution. On average, our Galactic cut removes only $\sim$11\% of the simulated subhalos, while 87\% of 2HWC unIDs are left out. Actually, in practice we cut a larger region $\abs{b}\leq 30^\circ$ in the N-body simulation in order to match the 2/3 sky coverage of HAWC, this way having a totally fair one-to-one comparison among observed and simulated sky.
\end{itemize}

After this selection, we are left with only three candidates in the 2HWC catalog, namely 2HWC J1309$-$054, 2HWC J1040$+$308 and 2HWC J0819$+$157. All three candidates are reported as very faint, i.e. near the source detection threshold of the catalog. In fact, \cite{Abeysekara2019} recently analyzed 760 days of HAWC data, i.e., approximately 50\% more integration time than the original 2HWC catalog, and found both 2HWC J0819$+$157 and 2HWC J1309$-$054 to be under the detection threshold. This behaviour of source flux versus integration time is not expected for actual sources but from spurious ones instead. Thus, in the following we only consider the remaining candidate for our analysis, 2HWC J1040$+$308. 
We note that a spectral analysis was already performed for this source in \cite{Abeysekara2019}, no preference for DM being found. Yet, more data are probably needed before being able to definitely reject this unID as a DM subhalo. Thus, we conservatively keep this source as the only surviving target in our list of potential subhalo candidates and set DM limits by assuming the existence of one single DM subhalo candidate in HAWC data.

\subsection{2HWC J1040$+$308 as a DM subhalo candidate}\label{sec:hawc_unids}
Before proceeding to set DM limits in the next section, we  would like to highlight here some properties of 2HWC J1040$+$308 that make this unID particularly appealing in a DM context, as such properties may be compatible with a DM origin for the observed emission:

\begin{itemize}
    \item \textbf{Spatial extension} -- Spatial extension has been hailed as a ``smoking gun'' for DM annihilation \cite{Bertoni+15, Bertoni+16, aguirre2019, Coronado-Blazquez2019_2}. Indeed, low-mass yet sufficiently close DM subhalos may be expected to appear as extended unID sources. 2HWC J1040$+$308 is found to be spatially extended in HAWC data with a radius of $0.5^\circ$ \cite{2HWC_paper}. This fact is even more notable for the case of very high latitude sources like this one ($b=61^\circ$), as sources at high latitudes are typically extragalactic and thus the majority of them are expected to be point-like. Indeed,it would be hard to explain all these features with a single astrophysical source -- AGNs would appear as point-like sources, while Galactic sources could appear as extended, yet this source's high latitude would imply a small distance and thus it would be surprising not to have a detection in any other wavelengths. On the other hand, as these unIDs correspond to very faint sources near the detection threshold, it is currently unclear whether they would appear as extended, even if being actual DM subhalos, as the DM annihilation flux profile decreases very rapidly with distance to the subhalo center.
    \\
    \item \textbf{Multi-wavelength search} -- As already mentioned, dark satellites are not massive enough to retain baryons and, as a result, they are not expected to shine at any wavelengths due to astrophysical processes. However, gamma-ray emission is expected to happen should these objects be composed of WIMP DM.\footnote{DM-induced emission may also be expected in radio due to synchrotron radiation from secondary particles \cite{Colafrancesco2006,Colafrancesco2007,Marchegiani2018,Cembranos2019}; yet the high density of radio sources would make any potential association probably very challenging.} A dedicated search at different wavelengths was performed for 2HWC J1040$+$308, with null results. In particular, a combined search in less energetic gamma rays with \textit{Fermi}-LAT and VERITAS was performed in \cite{VERITAS+HAWC+FERMI_paper}, with no detection. An additional search can be performed with the SSDC online tool,\footnote{\url{https://tools.ssdc.asi.it/}} where no significant emission at lower energies is found.
    \\
    \item \textbf{Heavy WIMP mass} -- It is interesting to note that a joint analysis of Fermi LAT \cite{fermi_instrument_paper}, VERITAS \cite{Park2015} and HAWC data was recently done in \cite{VERITAS+HAWC+FERMI_paper} and no gamma-ray emission was reported for 2HWC J1040$+$308 in the comparatively lower energy range covered by the LAT and VERITAS. This is so despite the fact that 2HWC J1040$+$308 exhibits a {\it hard} spectrum with a photon spectral index of $\Gamma=2.08\pm0.25$ in the HAWC energy range \cite{2HWC_paper}, which would in principle make a detection at low energies easier to realize. If interpreted in a DM scenario, these results suggest heavy, TeV WIMP masses
    as we would not expect a detection by Fermi LAT or VERITAS only in case of significantly high values of the WIMP mass, for which a DM spectrum beyond the range of sensitivity of these two instruments would be generated. Also, interestingly, this unID slightly prefers a fit to a ``Cutoff Power Law'' instead of a ``Simple Power Law'' \cite{Abeysekara2019}, which is a parametric form that better reproduces a typical DM annihilation spectrum \cite{Calore+17, Coronado_Blazquez2019}. Heavy WIMPs are well motivated \cite{Gammaldi2019,Cembranos2003,Cembranos2012} and, probably, favoured in the light of current DM constraints in the usual $\langle\sigma v\rangle$ (velocity-averaged annihilation cross section) vs. $m_{\chi}$ (WIMP mass) parameter space: IACTs are still far from being able to probe the \textit{thermal relic} cross section values \cite{Steigman+12} for large, TeV WIMP masses, while for much lighter, $\mathcal{O}$(GeV) masses there are already robust and strong constraints, e.g., \cite{dsphs_paper, Coronado_Blazquez2019}. 
\end{itemize}

All the above considerations make 2HWC J1040$+$308 an excellent candidate for being a DM subhalo. Yet, lacking a definitive answer for the moment we will proceed and will set constraints to DM annihilation using HAWC unIDs, just assuming that we cannot discard 2HWC J1040$+$308 as a DM subhalo.

\section{DM constraints}\label{sec:constraints}

The methodology we follow to set our DM constraints is explained in full detail in \cite{Coronado_Blazquez2019}. In short, these are computed according to i) the number of remaining HAWC unIDs as potential DM subhalo candidates, i.e. just one source (2HWC J1040$+$308), as discussed in Section \ref{sec:hawc_bigsection}; ii) the flux sensitivity of the instrument to DM annihilation; and iii) the J-factor predictions for the Galactic DM subhalo population as derived from N-body cosmological simulations. The annihilation cross section and WIMP mass $\langle\sigma v\rangle$ and $m_{\chi}$ are related by \cite{Coronado_Blazquez2019}:

\begin{equation}
\label{eq:master_formula}
\langle\sigma v\rangle=\frac{8\pi\cdot m_{\chi}^2\cdot F_{min}}{J\cdot \int_{E_{th}}^\infty\left(\frac{dN}{dE}\right)dE}
\end{equation}

\noindent where $F_{min}$ is the minimum flux needed for a point-source detection, $J$ is the astrophysical J-factor (see Eq. \ref{eq:j_factor}), and $\int_{E_{th}}^\infty\left(\frac{dN}{dE}\right)dE=N_{\gamma}$ is the integrated DM spectrum. This DM spectrum is obtained from the PPPC4 DM ID tables \cite{Cirelli+12}, including electroweak corrections, and integrated from $E_{th}=300$ GeV for a variety of annihilation channels. Hereafter, though, and for the sake of clarity, we will focus on $b\bar{b}$ and $\tau^+\tau^-$ channels, considering only pure annihilations (unity branching ratio).

\subsection{Minimum detection flux}\label{sec:fmin}
As mentioned previously, the minimum detection flux $(F_{min})$ is defined as the one required by the instrument to have a detection. More specifically, it is the source flux that is needed to reach $TS=25$ over the background, where TS is the Test Statistic, defined as,

\begin{equation}
    TS=-2~\mathrm{log}\left[\frac{\mathcal{L}(H_1)}{\mathcal{L}(H_0)}\right]
\end{equation}

\noindent $\mathcal{L}(H_1)$ and $\mathcal{L}(H_0)$ are, respectively, the likelihoods under the source and no source (null hypothesis) assumptions. Ideally, as done in \cite{Coronado_Blazquez2019}, one would need to account for the variations of this quantity across different instrumental setups (integration time, energy threshold, instrument response functions, etc.) and adopting different DM spectra.\footnote{This is so because, by default, this quantity is computed for a source spectrum with a power law index $\Gamma\sim2$, while the DM is better parametrized by a ``Power Law with SuperExponential Cutoff'' \cite{Coronado_Blazquez2019}, where the index and cutoff energy vary over the annihilation channel and WIMP mass.} Finally, the source latitude should also be taken into account, however this is a second-order effect once the Galactic plane is removed from the analysis, as it is the case.

Unfortunately, the VHE sky still lacks a proper model for the all-sky Galactic diffuse gamma-ray emission, which exists at lower energies \cite{fermi_galactic_diffuse_paper}. This, and the fact that we do not possess the HAWC instrumental response functions (IRFs), private for this Collaboration's internal use at the moment, precludes a precise characterization of the HAWC $F_{min}$. Instead, we will have to rely only on the differential sensitivity from figure 16 of \cite{fmin_hawc_paper}.\footnote{This sensitivity is computed for the Crab, which is located at DEC$\sim22^\circ$, while our source is at $\sim30^\circ$ instead. Fortunately, the HAWC sensitivity for both locations is expected to be very similar \cite{Abeysekara2013}.} At each energy, the $F_{min}$ can be obtained by dividing the differential sensitivity (in $\mathrm{TeV\cdot cm^{-2}\cdot s^{-1}}$) value by the log-central energy value in each considered bin (to obtain $\mathrm{ph\cdot cm^{-2}\cdot s^{-1}}$). In our case, we need the $F_{min}$ corresponding to a declination of $30.87^\circ$, which turns out to be very near the absolute minimum of the curve shown in figure 16 of \cite{fmin_hawc_paper}.\footnote{We note that, should we have had more than one unID, we would have computed a mean value after averaging the corresponding differential sensitivities for each source's declination.} Note, however, that the reported flux in \cite{fmin_hawc_paper} assumes a power law spectrum with a photon spectral index $\Gamma=2.63$ (Crab-like), while our DM subhalo candidate exhibits a considerably harder spectrum, $\Gamma=2.08$, and therefore it should be comparatively easier to detect. Also, the sensitivity is expected to scale as $\sim\sqrt{t}$ (assuming Poisson photon statistics), where $t$ is the exposure time. Thus, as the reported flux in \cite{2HWC_paper} was computed for 507 days of operation, the performance of HAWC would be underestimated for the 760 days of the current analysis.

With these limitations in mind, it is actually possible to compute a better estimate for $F_{min}$ than the one described above. We will define the improvement factor as the ratio between our estimation of $F_{min}$ and the official one. First, we take into account the improvement due to the photon spectral index -- a source with smaller index (harder spectrum), will be easier to detect than a source with larger index (softer spectrum). We can use the scaling relation reported in \cite{2HWC_paper}, i.e., the variation of $F_{min}$ with the source spectral index, for an energy value of 7 TeV, used as the pivot energy. By calibrating the improvement at this energy, we can extend it to all the considered energy range (300 GeV -- 100 TeV), taking into account the spectral shapes of both power laws. Then, we rescale the obtained $F_{min}$ by differences in exposure times, i.e., a factor $\sqrt{507/760}=0.82$.

\textbf{Also, as our unID is an extended source, with a reported extension of $0.5^\circ$, the HAWC sensitivity is expected to worsen with respect to that for a point-like source. Indeed, according to \cite{Abeysekara2013}, a $0.5^\circ$ source would require a flux 1.93 times larger to be detected at the same significance, independently of the assumed spectral index. As we are tailoring the sensitivity to our candidate, we will take this correction factor into account. As a result, the disagreement between our results and those in \cite{Abeysekara2019} is reduced to a factor $\sim25$.}

\begin{figure}[H]
\centering
\includegraphics[height=8cm]{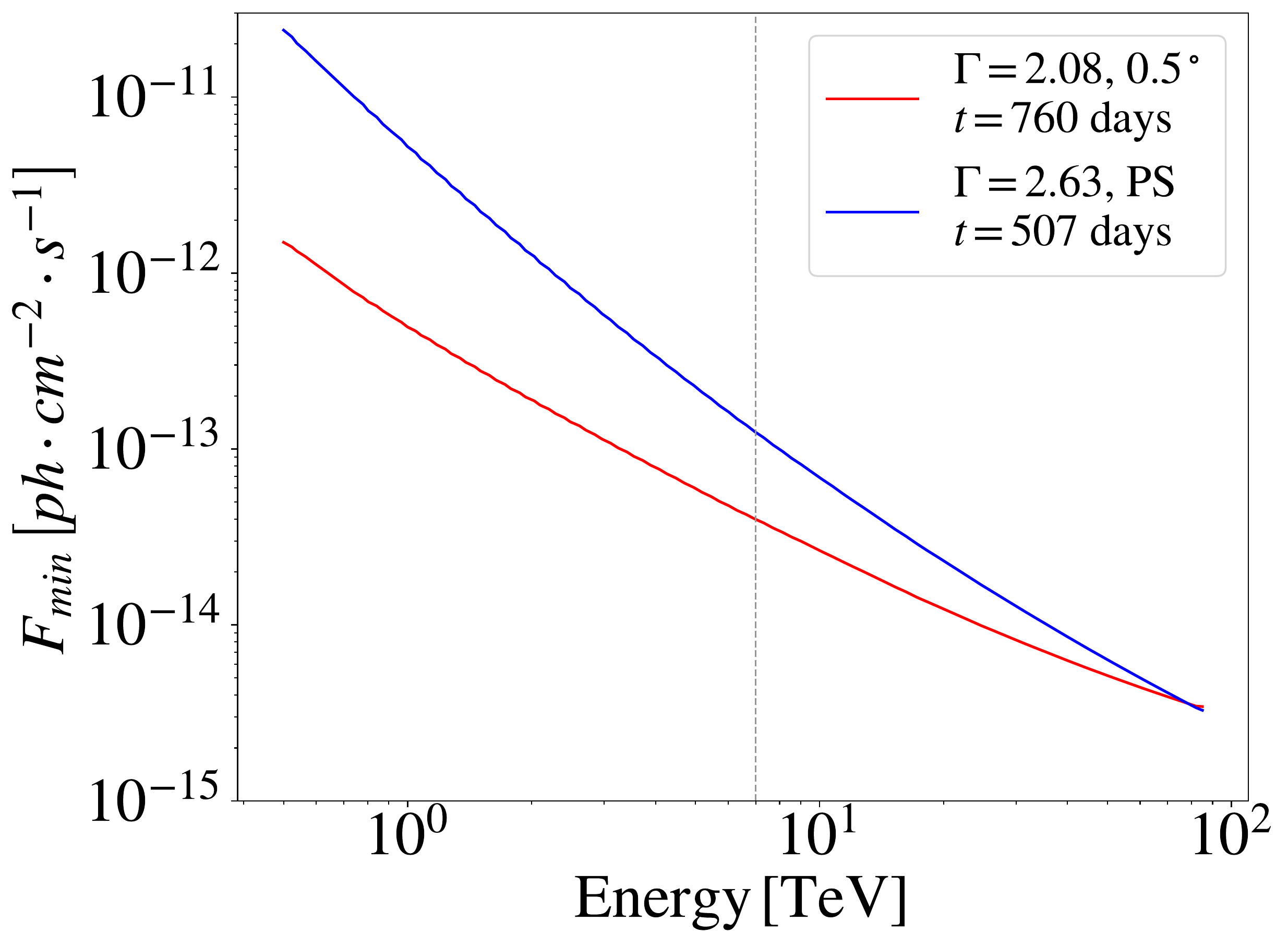}
\caption{HAWC minimum flux, $F_{min}$, needed to reach source detection ($TS=25$) at a sky declination equal to $30.87^\circ$. In blue, the $F_{min}$ as originally reported in \cite{fmin_hawc_paper} for the case of a point source described by a power law spectrum with index $\Gamma=2.63$ and 507 days of exposure. The red curve shows the $F_{min}$ for our DM subhalo candidate instead (with a spatial extension of $0.5^\circ$, $\Gamma=2.08$ and 760 days of exposure). This one is computed taking into account i) the improvement factor of $\sim5$ derived for this unID at a pivot energy of 7 TeV \cite{2HWC_paper}, marked in the figure as a dashed grey vertical line; ii) the difference in exposure times, which makes the $F_{min}$ improve by a factor $\sqrt{507/760}=0.82$; and iii) a factor $\sim$2 worsening due to the extension of the source, according to \cite{Abeysekara2013}. See text for further details on these computations. The overall $F_{min}$ improvement is roughly an order of magnitude at low energies, while at the highest energies there is no improvement at all.}
\label{fig:fmin}
\end{figure}

A comparison between both the originally reported $F_{min}$ in \cite{fmin_hawc_paper} ($\Gamma=2.63$, $t=507$ days, point-like) and the ``improved'' $F_{min}$ here described for our candidate ($\Gamma=2.08$, $t=760$ days, $0.5^\circ$) is shown in Figure \ref{fig:fmin} as the blue and red curves, respectively.  The latter is the one that will be used to set our DM constraints in Section \ref{sec:results}.\footnote{The spectral index is reported to have an uncertainty $\Gamma=2.08\pm0.25$. We checked that this uncertainty translates into a factor $\sim$4 uncertainty in $F_{min}$ at most (factor 2 when compared to the benchmark $\Gamma=2.08$).}

\subsection{J-factor}\label{sec:jfactor}
The last ingredient needed to set the DM constraints is the J-factor. In the case of unID-based DM searches, we conservatively assume that all $N$ DM subhalo candidates in our unID list ($N=1$ in this work) are in fact DM subhalos \cite{Calore+17,Coronado_Blazquez2019}. Then, we require consistency of this supposedly observed number of DM subhalos with that obtained from N-body simulations, by assuming that the J-factor of the $N^{th}$ (1$^{st}$, in this work) most brilliant DM subhalo in our simulation sets the border between the population of detected and non-detectable subhalos.

\begin{figure}[H]
\centering
\includegraphics[height=9cm]{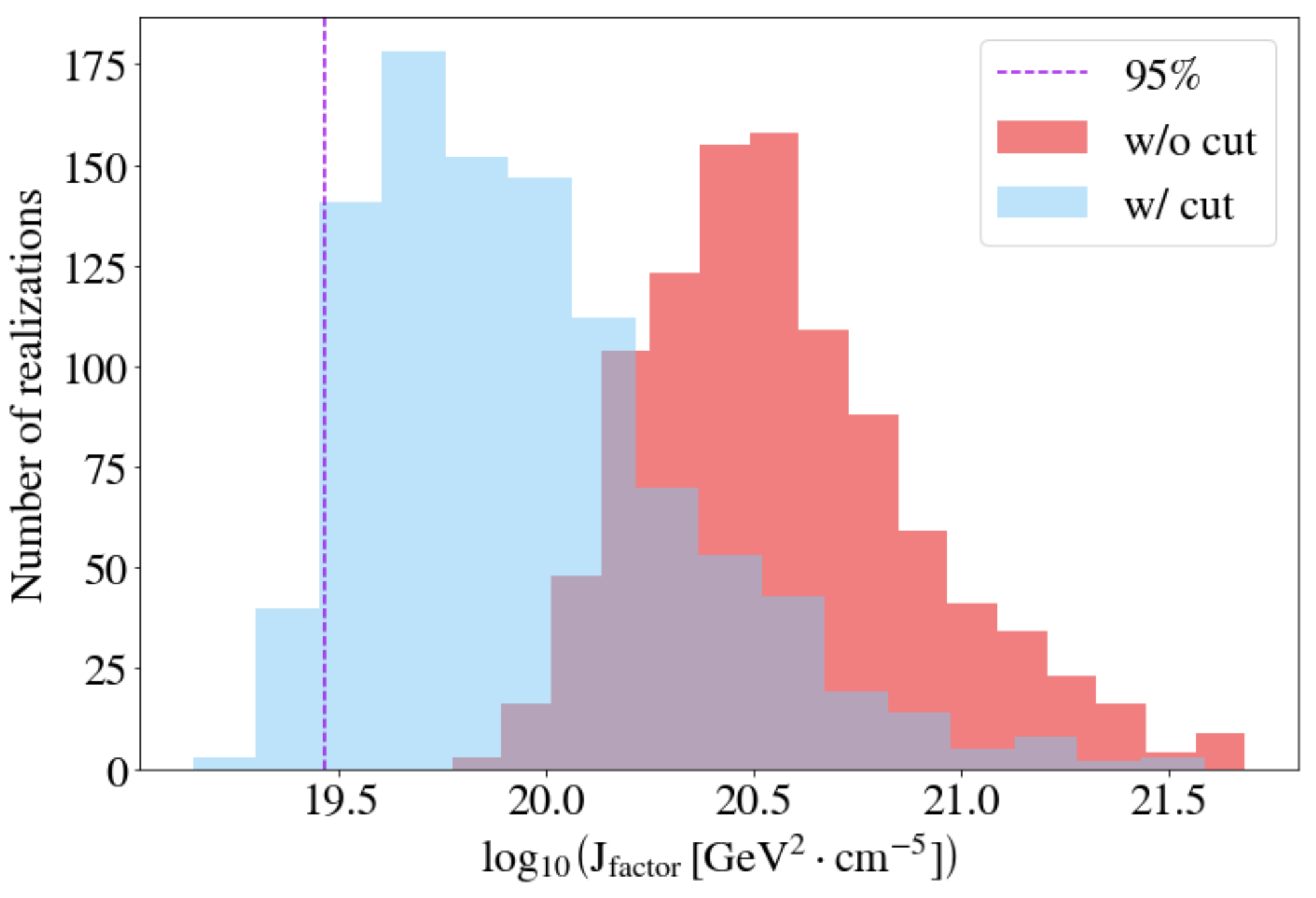}
\caption{J-factor distributions for the brightest subhalo across 1000 realizations of the DM subhalo population as derived in \cite{Coronado_Blazquez2019} from the Via Lactea II N-body simulation. Red and blue histograms show, respectively, the values before and after applying our selection cuts on the simulated DM subhalos, namely a mass cut ($M\leq10^7\mathrm{M_\odot}$) and a coordinate cut ($|b|>30^\circ$). The dashed, purple vertical line marks the value of the J-factor that will be used to set the 95\% C.L. DM limits in Section \ref{sec:constraints}, and it is defined as the one above which 95\% of the J-factor distribution is contained.}
\label{fig:jfact_dist}
\end{figure}

As input, we use the N-body simulation work by \cite{aguirre2019}, specifically designed to assess the relevance of low-mass subhalos for this kind of studies. The authors of \cite{aguirre2019} {\it repopulate} the original Via Lactea II DM-only N-body cosmological simulation of a Milky-Way-size halo \cite{vlii_paper} with subhalos well below its resolution limit by applying bootstrapping techniques and semi-analytical extrapolations of the relevant quantities. In particular, the subhalo mass function is extended down to $10^3$M\textsubscript{\(\odot\)}, the subhalo distribution within the host halo is assumed to be similar to that of the resolved subhalos in the parent simulation, and a state-of-the-art concentration model \cite{Moline+17} is used to model the subhalo structural properties and, ultimately, to compute the J-factors. Actually, hundreds of realizations of the parent simulation are performed that allow to derive statistical meaningful J-factor results. More precisely, as done in \cite{Coronado_Blazquez2019}, in order to derive 95\% C.L. upper limits on the DM annihilation cross section, we obtain the distribution of J-factor values for the $N^{th}$ subhalo under consideration and then as reference J-factor the one above which 95\% of the J-factor distribution is contained; see Figure \ref{fig:jfact_dist}.

In this particular case, we cut out from the simulation results the Galactic region $\abs{b}\leq 10^\circ$ in order to match the cut that we applied on the data to avoid the Galactic plane contamination (see Section \ref{sec:target_selection}). We also take into account that HAWC, unlike Fermi-LAT, does not achieve a uniform exposure for the whole sky but only for $\sim2/3$ of it. Specifically, to mimic this effect in the simulation side, we cut out the $\abs{b}\leq30^\circ$ region (i.e., we only keep 2/3 of the whole sky) in every realization. Finally, as we are interested in dark satellites alone, only M$\leq10^7$M\textsubscript{\(\odot\)} subhalos (i.e., not massive enough to retain baryons) are considered. We checked that the resulting distribution looks isotropic as seen from the Earth. The result of applying and of not applying cuts on the J-factor distributions for the most brilliant subhalo in the simulations is shown in Figure \ref{fig:jfact_dist}. The value used to set the constraints, above which 95\% of the distribution is contained, is $\mathrm{log}_{10}(J)=19.47$, a factor 4.26 smaller than the value before cuts.\footnote{The J-factor of our brightest subhalo is comparable to the one typically quoted for dwarf spheroidal satellite galaxies (see e.g. \cite{Albert2017}), which are expected to be well above the extragalactic isotropic DM-induced gamma ray background. On the other hand, the value of the DM-induced Galactic diffuse emission is comparable to the isotropic contribution at high latitudes \cite{isotropic_fermi15}, and therefore also negligible compared to the brightest subhalo J-factor. Finally, the boost due to Galactic unresolved substructure contributing to the diffuse emission in the line of sight of this unID is expected to be at the level of few percent and, thus, not relevant here. The boost due to substructures is only particularly important when integrating the subhalo signal for the whole host halo; see e.g. the discussions and results in \cite{Moline+17}.}

\subsection{Results}
\label{sec:results}
\begin{figure}[!ht]
\centering
\includegraphics[height=5.75cm]{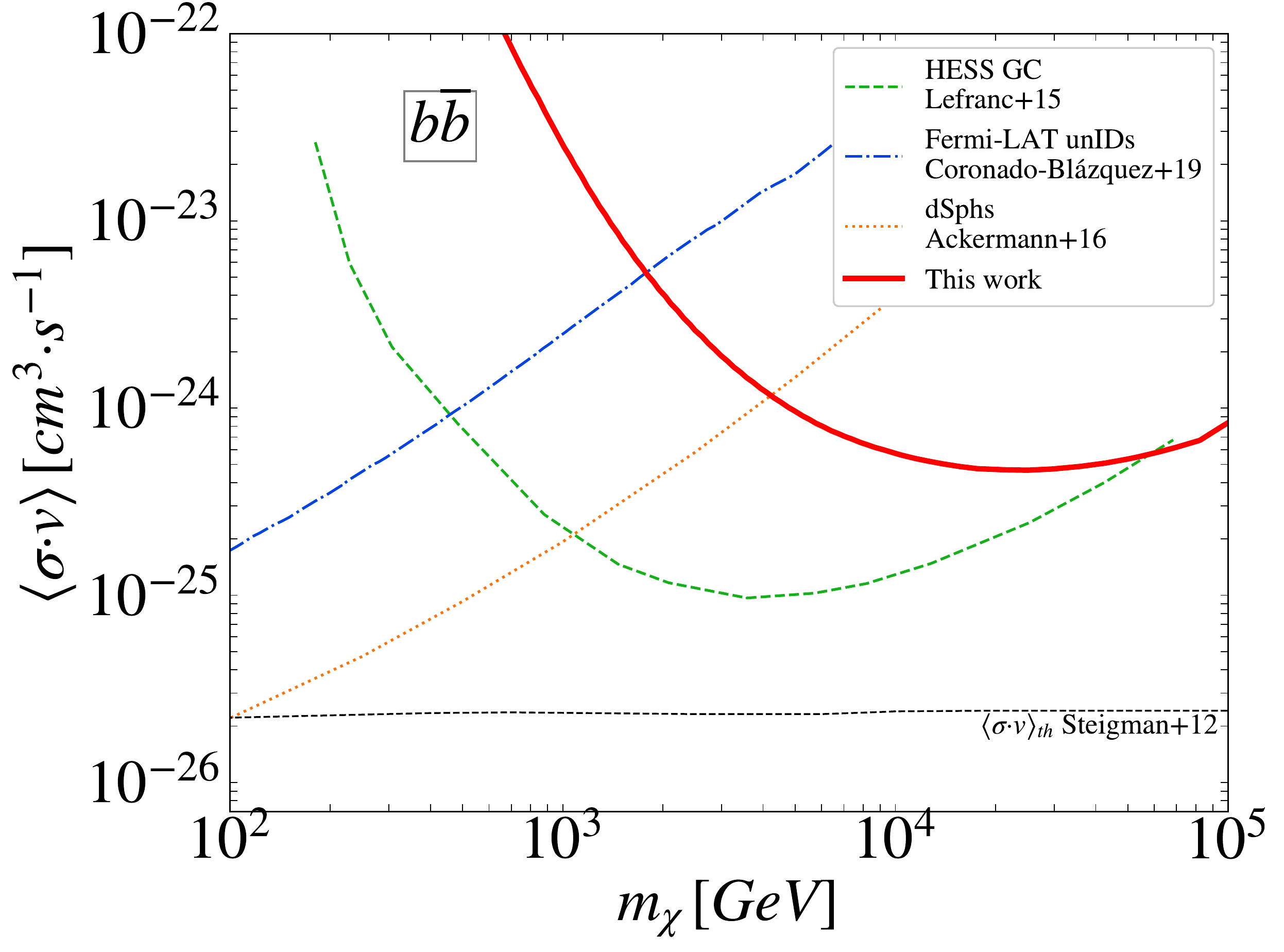}
\hfill
\includegraphics[height=5.75cm]{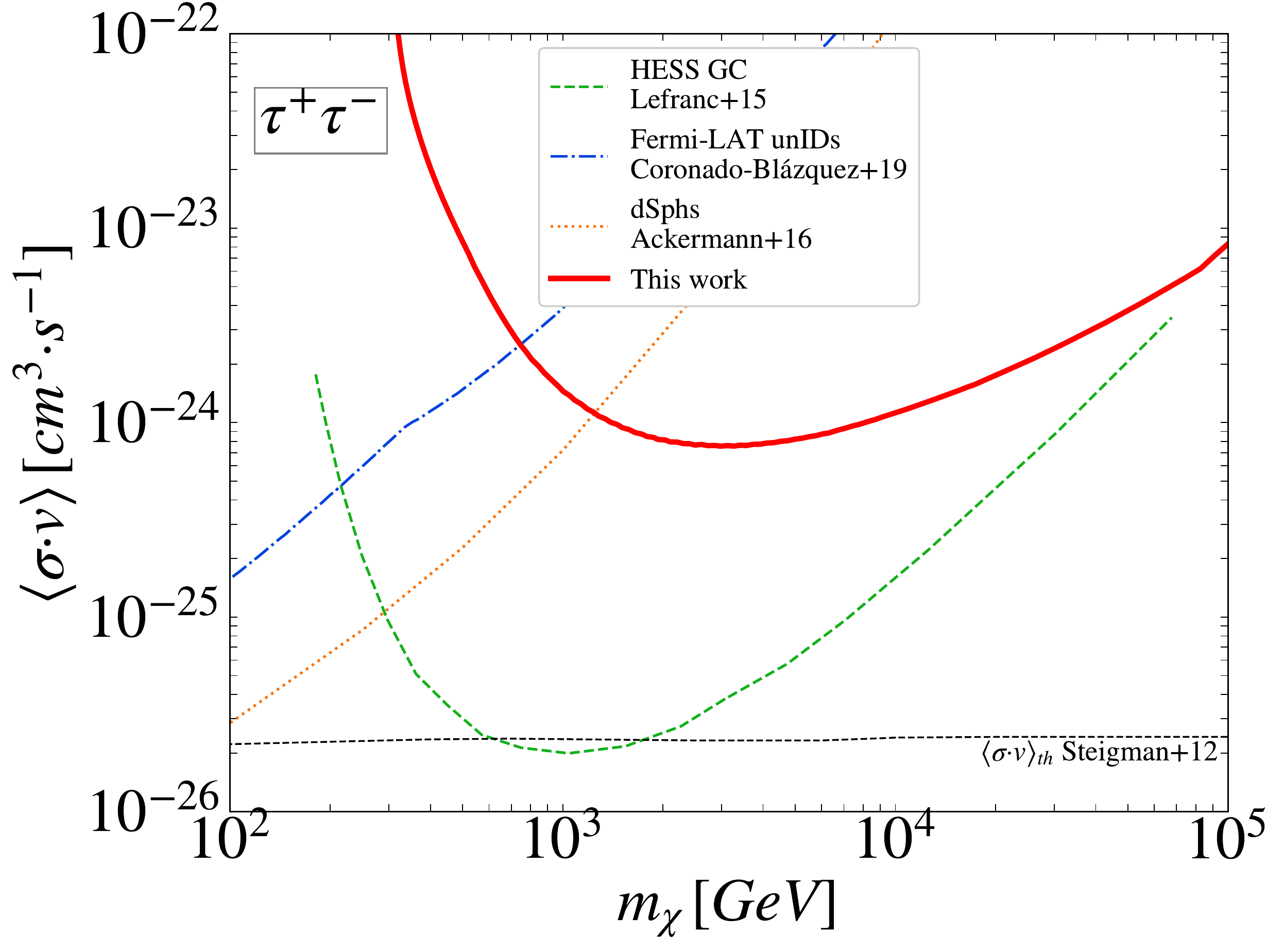}
\caption{95\% upper limits on the DM annihilation cross section as derived from HAWC unIDs and predictions from N-body cosmological simulations. Indeed, these constraints use 2HWC J1040$+$308 alone, i.e., the only HAWC unID located at $\abs{b}\geq 10^\circ$ and surviving our selection criteria in Section \ref{sec:target_selection}. Left (right) panel shows the 95\% C.L. upper limits for the $b\bar{b}$ ($\tau^+\tau^-$) annihilation channel. For comparison, we also show as a blue, dot-dashed line the DM constraints obtained from Fermi-LAT unIDs using the same methodology \cite{Coronado-Blazquez2019_2}. Limits from the observation of the Galactic center region by H.E.S.S., i.e., the best DM constraints achieved from IACT observations at present, are included in both panels as well as a green, dashed line \cite{Hess_Lefranc+15}. Finally, a dotted, orange line shows the constraints from \textit{Fermi}-LAT observations of dSphs \cite{dsphs_paper}. Note that we did not include in this Figure the results in \cite{Abeysekara2019}, as a one-to-one comparison would be misleading given the fact that the methodology and underlying assumptions in both works are significantly different; see discussion in Section \ref{sec:constraints} for details.}
\label{fig:constraints_bb_tautau}
\end{figure}

Once all involved quantities in Equation \ref{eq:master_formula} have been properly characterized, we can set constraints on the DM parameter space at 95\% confidence level. These are plotted in figure \ref{fig:constraints_bb_tautau} for the $b\bar{b}$ and $\tau^+\tau^-$ annihilation channels. We remind, once again, that these limits adopt a single HAWC unID as being a potential DM subhalo (2HWC J1040$+$308). 

The obtained 95\% upper limits to the DM annihilation cross section are most stringent for masses of $\sim$20 (2) TeV for the $b\bar{b}$ ($\tau^+\tau^-$) annihilation channel, reaching cross section values around $5\cdot10^{-25}$ ($8\cdot10^{-25}$) $cm^3s^{-1}$. This is roughly one order of magnitude far from the thermal relic cross section value, yet they are competitive with today's IACTs best constraints (derived from H.E.S.S. observations of the Galactic center \cite{Hess_Lefranc+15}) for very heavy WIMPs annihilating to $b\bar{b}$ above $\sim$50 TeV. It is worth emphasizing that our DM constraints are independent and complementary to the ones obtained for other astrophysical targets and by means of different analysis techniques.

We must note here that we observe a mismatch between the DM constraints obtained in this work and the ones shown in \cite{Abeysekara2019}. This disagreement, that reaches a factor $\sim25$ for $\tau^+\tau^-$, is hard to understand in detail but it is probably due to several factors. In particular, our $F_{min}$ is expected to be a fair approximation while \cite{Abeysekara2019} adopts the actual sensitivity of HAWC to a DM spectrum (only computable with the IRFs). In that work, the structural properties of subhalos were described according to \cite{Sanchez-Conde2014}, which is well-suited for field halos, not subhalos. Indeed, we use \cite{Moline+17}, where subhalos exhibit concentration values a factor 2-3 larger than field halos of the same mass. As the J-factor is roughly proportional to the third power of the concentration, this correction would thus yield J-factors about a factor 8-9 times larger, leading to better constraints. Other possible systematics, such as those coming from the Galactic diffuse emission model and the observation strategy adopted in \cite{Abeysekara2019}, may turn out to be particularly relevant. Also, we note that they show cross-section values needed for detection, while we here show $95\%$ upper limits. All in all, a precise one-to-one comparison with the results in \cite{Abeysekara2019} is not possible, as the methodology and underlying assumptions are different.

For the sake of comparison, we also show in figure \ref{fig:constraints_bb_tautau} the constraints found by \cite{Coronado_Blazquez2019} following a similar methodology, but using unIDs in \textit{Fermi}-LAT catalogs instead. Both these low and high energy unID-based DM limits meet at roughly 2 (0.7) TeV for $b\bar{b}$ ($\tau^+\tau^-$) annihilation channel.

Finally, it is important to note that our DM limits are subject to potentially large uncertainties. In addition to the already mentioned caveats in the computation of the minimum detection flux (see Section \ref{sec:fmin}),\footnote{Namely, the lack of both the HAWC IRFs and a Galactic diffuse TeV emission model.} there exist important theoretical uncertainties in the N-body simulation predictions, such as the survival probability of the lightest subhalos near the Galactic center \cite{vandenbosch, garrison-kimmel, Errani2019},\footnote{\textbf{We note that tidal stripping is implicitly included in our simulations. Yet, we note that the mentioned works claim that the disruption of a large fraction of low-mass subhalos in simulations may be artificial and numerical in origin -- if this was indeed the case, the distribution of J-factors of the entire subhalo population would surely reach larger values, as we would expect more and closer-to-Earth subhalos. Therefore, our results are conservative, as the DM limits would become even more stringent for resilient subhalos.}} the precise subhalo structural properties, the impact of baryons on the subhalo population \cite{Kelley2019,Calore+17} or the value of the minimum subhalo mass that is adopted to separate between visible and dark satellites \cite{Sawala2017}. These and other issues that affect the Galactic subhalo population will be addressed elsewhere. Also, for WIMP masses above $\sim10$ TeV, the theoretical spectra of \cite{Cirelli+12} should include higher-order EW corrections, although these are expected to be potentially relevant only for leptonic channels such as $e^+e^-$ and $\mu^+\mu^-$.

\section{Summary and conclusions}
\label{sec:summary}
In this work, we have studied the HAWC unIDs reported in the 2HWC catalog to search for potential DM subhalos. HAWC proves to be the best VHE gamma-ray observatory to perform this kind of search at present, as it is the only one that surveys a significant portion of the sky in this energy window.

Starting from a total of 23 unIDs in the 2HWC, we apply several selection criteria in Section \ref{sec:target_selection} based on expected DM signal and subhalo properties. More precisely, at first we only select those HAWC unIDs located at Galactic latitudes $\abs{b}\geq10^\circ$ to avoid contamination from astrophysical sources along the Galactic plane. 87\% of the 2HWC unIDs do not pass this cut, i.e. we are left with only three unIDs. We then consider these three unIDs as DM subhalo candidates, yet two of them do not even reach the flux detection threshold when adding $\sim$250 more days of data \cite{Abeysekara2019}. Therefore, only one unID, labeled as 2HWC J1040+308, remains as a potential DM subhalo after our filtering procedure. Indeed, the measured flux of this unID grows as expected when considering more integration time, which reinforces its actual existence. 

When scrutinized in further detail, 2HWC J1040+308 turns out to be specially interesting in a DM context, as it is reported in the 2HWC catalog as spatially extended, it exhibits a hard spectrum and does not possess visible counterparts neither at other wavelengths nor for other gamma-ray observatories operating at lower energies such as \textit{Fermi}-LAT and VERITAS. The combination of the latter with the hard source spectrum might point towards particularly heavy WIMP masses, if the gamma-ray emission was actually produced by DM annihilation.

While waiting for more data that can help to shed further light on the true nature of 2HWC J1040+308, we proceed in Section \ref{sec:constraints} and set limits on the DM annihilation cross section by conservatively assuming that this source is actually a subhalo. To do so, we follow the methodology in \cite{Coronado_Blazquez2019} and compare the HAWC unID observations with predictions for the Galactic subhalo population as derived from N-body cosmological simulations of a galaxy like our own. In particular, we follow the simulation work in \cite{Coronado_Blazquez2019,aguirre2019}, where the VL-II DM-only simulation was repopulated to include subhalos with masses down to $10^3$M\textsubscript{\(\odot\)}. This work provides the expected subhalo J-factors, which we combine with our own estimate of the HAWC instrumental sensitivity to DM subhalos in order to set 95\% C.L. upper limits in the $\langle\sigma v\rangle$-$m_{\chi}$ parameter space. 

Our DM limits are most stringent for masses of $\sim$20 (2) TeV for the $b\bar{b}$ ($\tau^+\tau^-$) annihilation channel, reaching cross section values around $5\cdot10^{-25}$ ($8\cdot10^{-25}$) $cm^3s^{-1}$. Although roughly one order of magnitude far from the thermal relic cross section value, our limits are competitive with today's IACTs best constraints \cite{Hess_Lefranc+15}) above $\sim 50$ TeV for $b\bar{b}$ annihilation channel. Also, these constraints are independent and complementary, yet competitive, to those obtained for other astrophysical targets adopting diverse analysis techniques.

The analysis presented in this work is subject to some potentially important uncertainties. In particular, a proper determination of the minimum HAWC detection flux taking into account the specific instrumental setup used in the observations, as well as the DM spectrum, annihilation channel and WIMP mass, would be required. The HAWC IRFs are nevertheless not public, thus this study is beyond the scope of the current paper. Instead, we used public information in \cite{2HWC_paper,fmin_hawc_paper} to come up with a fair estimate of the HAWC sensitivity to DM subhalos (Section \ref{sec:fmin}).  
Another source of uncertainty is the lack of a proper Galactic diffuse gamma-ray background model, which prevents us from including it in our estimate of the HAWC sensitivity to DM. Yet, as in the end only one unID survives our cuts, this single unID is located at high Galactic latitudes (where the diffuse emission is expected to be subdominant) and exhibits a hard spectrum well fitted by a power law, we expect this uncertainty from the Galactic diffuse model to be a second order effect. 
Finally, there are also theoretical uncertainties hidden behind the adopted N-body simulation work. In particular, the survival probability of the smallest subhalos near the Galactic center --  thus subject to strong tidal forces -- is a matter of fierce debate in the community at present, with different works \cite{vandenbosch, garrison-kimmel, Errani2019} providing diverse answers. In addition, the impact of baryons on the subhalo population was not considered here. Baryons may induce a reduction of the number of subhalos near the Sun's position, which would worsen our DM constraints probably by a factor of a few \cite{Calore+17,Kelley2019}.

Looking into the future, more data, preferably not only in the VHE regime, are needed in order to elucidate the true nature of 2HWC J1040+308, i.e., the only HAWC unID that survives our selection cuts and, indeed, a promising DM subhalo candidate. Related to this, we note that the simplest and most robust way to improve our DM constraints is to also associate 2HWC J1040+308 to a known astrophysical source. In fact, a simple detection of variability or multi-wavelength emission would be enough to discard it as a DM subhalo (see several examples of unID rejections in \cite{Coronado_Blazquez2019}). 
As HAWC accumulates exposure time, also its minimum detectable flux will be lowered, thus allowing for better DM limits for a fixed number of unIDs. New HAWC observations and additional exposure time will surely bring up new unIDs though, thus in the end the resulting DM limits may improve or worsen depending on the number of unIDs that will be potential DM subhalos \cite{Calore+17,Coronado_Blazquez2019}. 

Finally, the upcoming Cherenkov Telescope Array (CTA) \cite{CTA_science_paper}, with its superb instrumental capabilities, enhanced sensitivity to DM, improved angular resolution to pinpoint source extension, and already planned surveys of large portions of the VHE sky with unprecedented sensitivity, should be able to set the most competitive DM limits in the TeV energy range by means of the unID-based method used in this work. Alternatively, it is also possible that CTA may bring the first robust DM subhalo discovery.

\acknowledgments{The authors would like to thank Hugo Alberto Ayala Solares, Viviana Gammaldi and Colas Rivi\`{e}re for valuable help and feedback. JCB and MASC are supported by the {\it Atracci\'on de Talento} contract no. 2016-T1/TIC-1542 granted by the Comunidad de Madrid in Spain, by the Spanish Agencia Estatal de Investigaci\'{o}n through the grants PGC2018-095161-B-I00, IFT Centro de Excelencia Severo Ochoa SEV-2016-0597, and Red Consolider MultiDark FPA2017-90566-REDC.}

\reftitle{References}
\externalbibliography{yes}
\bibliographystyle{mdpi.bst}
\bibliography{References.bib}

\end{document}